# Ultrafast all-optical switching via coherent modulation of metamaterial absorption

Xu Fang[1], Ming Lun Tseng[2,3], Jun-Yu Ou[1], Kevin F. MacDonald[1], Din Ping Tsai[2,3,4], and Nikolay I. Zheludev[1,5]

[1]*Optoelectronics Research Centre and Centre for Photonic Metamaterials, University of Southampton, Southampton SO17 1BJ, UK*
[2]*Graduate Institute of Applied Physics, National Taiwan University, Taipei 106, Taiwan*
[3]*Department of Physics, National Taiwan University, Taipei 106, Taiwan*
[4]*Research Center for Applied Sciences, Academia Sinica, Taipei 115, Taiwan*
[5]*Centre for Disruptive Photonic Technologies, Nanyang Technological University, Singapore 637371, Singapore*

We report on the demonstration of a femtosecond all-optical modulator providing, without nonlinearity and therefore at arbitrarily low intensity, ultrafast light-by-light control. The device engages the coherent interaction of optical waves on a metamaterial nanostructure only 30 nm thick to efficiently control absorption of near-infrared (750-1040 nm) femtosecond pulses, providing switching contrast ratios approaching 3:1 with a modulation bandwidth in excess of 2 THz. The functional paradigm illustrated here opens the path to a family of novel meta-devices for ultrafast optical data processing in coherent networks.

A move from electronic to all-optical signal switching/processing in telecommunications and data networks has long been regarded as desirable to increase bit rates and reduce latency while also reducing energy consumption and simplifying network structure.[1-5] Indeed, with the fundamental capacity limits of existing infrastructure now being reached,[6] future architectures will require a new generation of highly integrated, ultrafast devices capable of functions such as all-optical switching and mode (de)multiplexing, which in turn will rely on the development of advanced materials and metamaterials with a range of novel properties and functionalities.[7-12] Coherent optical networks now achieve 100 gigabit per second data rates by encoding information not only in the binary presence or absence of light but in the amplitude, phase and polarization of signals. They provide for the application of advanced detection and digital signal processing algorithms, and present unique opportunities for implementing novel signal control mechanisms. Indeed, coherent networks provide an ideal environment in which to exploit the recently demonstrated phenomenon of coherently controlled metamaterial transparency/absorption: the interference of two continuous, counter-propagating coherent beams on a photonic metamaterial of subwavelength thickness can, depending on their mutual intensity and phase and on their polarization, either entirely eliminate Joule losses in the metallic nanostructure or lead to the total absorption of all incident light.[13] Here, we illustrate the applications potential of this effect by demonstrating a four-port, ultrafast, all-optical coherent 'meta-device' modulator providing, without nonlinearity and therefore at arbitrarily low intensity, light-by-light control of femtosecond pulses with 2.2 THz bandwidth. (It should be emphasized that this mechanism for 'coherently control' of light-matter interactions is, descriptive terminology aside, distinctly different from recently

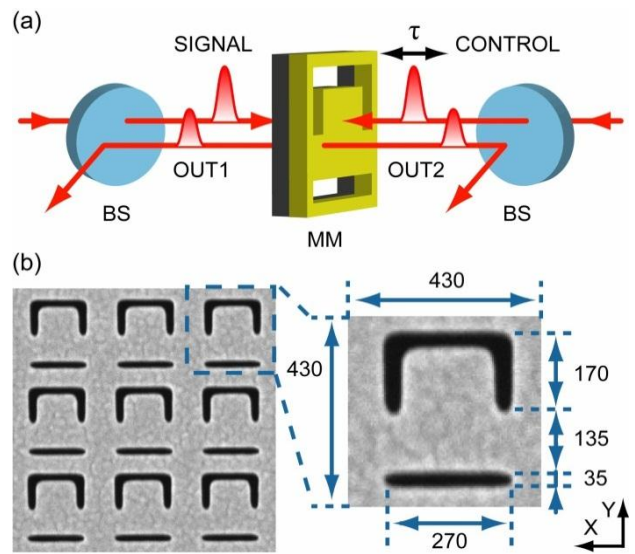

FIG. 1: Ultrafast coherent metamaterial modulator. (a) The functional element, a metamaterial (MM, gold on silicon nitride) nanostructure of sub-wavelength thickness, is illuminated by counter-propagating coherent femtosecond input pulses (nominally SIGNAL and CONTROL); modulator output comprises the sum of transmitted and reflected beams from both directions (OUT1 + OUT2). The relative time delay τ between the two inputs is tuned by using a combined mechanical motor/piezoelectric translation stage. Separation of inputs and outputs is realized using pellicle beamsplitters (BS). (b) Scanning electron microscope images showing a section of an experimental sample and dimensions of the nanostructure.

reported concepts based on phase modulation of single ultrashort excitation pulses.[14,15])

The experimental modulator's input/output beam configuration is illustrated schematically in Fig. 1(a). Light generated by a mode-locked Ti:sapphire laser (with an output pulse duration of order 130 fs, a central wavelength tunable from 750 to 1040 nm and a spectral full-width half-maximum of 10-20 nm) was intensity modulated by a mechanical chopper and then divided by a pellicle into two beams (nominally 'signal' and 'control'), constituting the two optical inputs to the coherent modulator. A combined mechanical motor/piezoelectric translation stage located in the control beam path sets/tunes the relative time delay and thus the relative phase difference between incident pulses. The beams were focused at normal incidence from opposite sides onto a metamaterial sample using plano-convex lenses. Their average powers at the sample position were balanced (at each measurement wavelength) using a variable neutral density filter, and maintained at a level below 1 mW per beam to exclude undesired opto-thermal or nonlinear effects. The two output beams (transmitted and reflected from both sides of the metamaterial) were directed via pellicle beamsplitters to a pair of identical photodiodes, the signals from which were monitored using lock-in amplifiers referenced to the input beam chopping frequency of 1.6 kHz. Signal levels were calibrated at each measurement wavelength to eliminate the influence of optical components, e.g. variations in the energy splitting ratio of the pellicles.

Ideally, the functional element of the coherent modulator should be a vanishingly thin film which, at any given wavelength within the operational range, absorbs half of the energy of a single incident beam.[13] For fundamental reasons an infinitely thin film cannot absorb more than 50% of an incident beam,[16,17] and practically this level is difficult if not impossible to achieve in an unstructured thin (sub-wavelength) films. However, through metamaterial nanostructuring one can achieve, by design, a workable balance among absorption, reflection and transmission characteristics at any designated wavelength. In the present case, the metamaterial sample was fabricated by depositing a 30 nm gold film on a low-stress silicon nitride membrane (50 nm thick) by thermal evaporation. An array of asymmetric split-ring (ASR) slits, covering a total area of approximately 50 μm × 50 μm, was cut through both the gold and silicon nitride layers by focused ion beam (FIB) milling (from the silicon nitride side to minimize damage to the gold layer). The high fabrication quality and uniformity of the sample is confirmed by scanning electron microscope images such as shown in Fig. 1(b). With a unit cell size of 430 nm the sample is non-diffracting throughout the wavelength range of interest in the present study. Detailed structural dimensions, as used in FIB pattern design and in 3D computational modeling of the metamaterial (using the COMSOL Multiphysics finite-element solver) are presented in Fig. 1(b). In all experiments and numerical simulations presented here, incident light is polarized in the *Y* direction (as defined in Fig. 1(b)).

The metamaterial's single-beam reflection, transmission and thereby absorption spectra were obtained for both surface-normal illumination directions using a microspectrophotometer (Figs. 2(a) and 2(b)), with reflection and transmission levels normalized against those of a silver mirror and air respectively. A resonance, appearing as a dip in reflection and a peak in absorption and transmission, is observed at ~900 nm in both panels and is attributed to the trapped electromagnetic mode of the ASR nanostructure.[18]

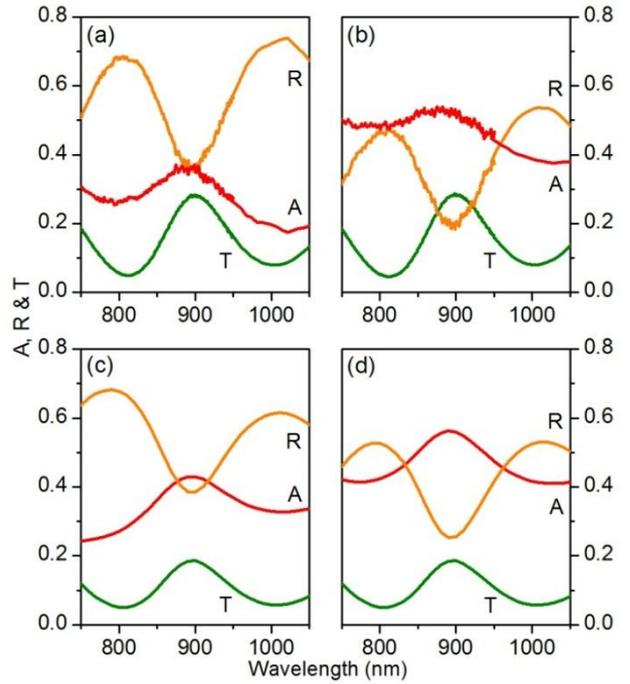

FIG. 2: Metamaterial optical properties: (a, b) Measured metamaterial reflection R, transmission T and absorption A spectra for single-beam illumination from (a) the gold side and (b) the silicon nitride side of the sample. (c, d) Corresponding numerically simulated spectra for illumination from (c) the gold side and (d) the silicon nitride side.

Corresponding numerically simulated spectra are presented in Figs. 2(c) and 2(d). The computational model assumes silicon nitride to be lossless with a relative permittivity equal to 4 across the near-infrared wavelength range of interest to the present study,[19] and derives the permittivity of gold from a Drude-Lorentz model using a Drude damping term three times that of bulk gold to account for thin film surface roughness.[20,21] These parameters produce a very good correlation with experimental data, with remaining discrepancies being attributed primarily to manufacturing imperfections in the sample. In both theory and practice, reflection and absorption are seen to depend on illumination direction while, in accordance with requirement for a linear, reciprocal system, the transmission does not.

Figure 3 shows, for a wavelength of 900 nm, the total output of the metamaterial modulator (the sum of the two photodiode signals) as a function of the mutual time delay (temporal phase offset) between pulses arriving at the sample via the two input paths. At large positive and negative delays the output of the system is constant at a level corresponding to the total *incoherent* absorption of the two input beams by the metamaterial (see Fig. 3(a)) – the arrival of a pulse at the metamaterial from one direction is well separated in time from that of the corresponding pulse from the opposite direction and they therefore interact independently with the nanostructure. In comparison, where there is a temporal overlap between counter-propagating pulses in the plane of the metamaterial, (delay times between approximately -300 and +300 fs) the system oscillates between regimes of coherently enhanced absorption (resulting in reduced signal output) and coherent transparency (i.e. suppressed absorption, giving increased signal output[13]).

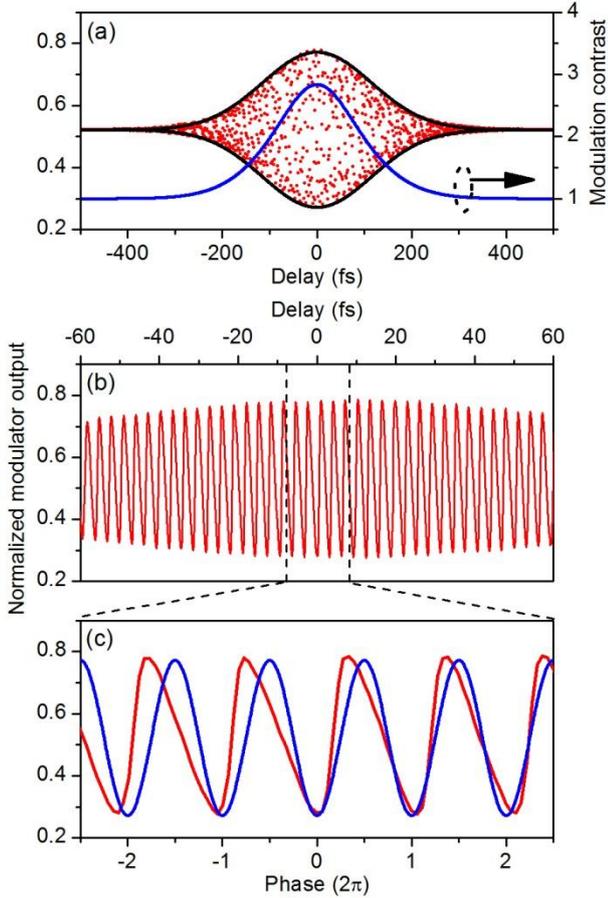

FIG. 3: Femtosecond pulse modulation using a plasmonic metamaterial: (a) Dependence of metamaterial coherent modulator output power (relative to total input) on the temporal delay between counter-propagating input pulses at a wavelength of 900 nm. Experimental data (red dots) are overlaid with analytical envelope curves given by Eq. (1) (black line) and the corresponding modulation contrast (blue line). (b) Data for the ±60 fs range obtained with 0.2 fs temporal resolution using the piezoelectric translation stage. (c) Detail of the central region of panel (b) with time delay converted to mutual optical phase, overlaid with an analytical curve given by Eq. (2) (blue line).

The time-dependent electric field at the metamaterial plane can be expressed as the sum of the two incident pulse fields $E(t) + E(t-\tau)$, where $\tau$ is the time delay between the two. The level of coherent absorption achieved in response to this applied field is then equal to

$$A'_1 \left[1 + \frac{Re \int_{-\infty}^{+\infty} E(t)E^*(t-\tau)dt}{\int_{-\infty}^{+\infty} E(t)E^*(t)dt}\right] + A'_2 \qquad (1)$$

where $A'_1$ is the incoherent absorption, and $A'_2$ encompasses residual losses (such as incoherent scattering due to surface roughness) which are intrinsic to single-beam absorption values derived from measurements of specular reflection and transmission but which do not contribute to coherent absorption. The level observed when $\tau$ is significantly larger than the pulse duration, e.g. at ±500 fs delay in Fig. 3(a), is equal to $A'_1 + A'_2$. This analysis assumes that the metamaterial is a vanishingly thin absorber with a linear response - a good approximation because absorption occurs only within the gold layer (which has a thickness an order of magnitude smaller than the laser wavelength) and because the dephasing time in plasmonic nanostructures (a few tens of femtoseconds[22,23]) is an order of magnitude smaller than the pulse duration. (Note that under zero-thickness, perfectly-smooth assumptions $A'_2 = 0$ and $A'_1$ must be $\leq 0.5$.) An infinite interval is taken for the integral because the detector response time is much greater than the pulse duration, i.e. they do not resolve power fluctuation inside single pulses. The field autocorrelation function[24] in Eq. (1) describes all the essential features of the metamaterial response, and therefore coherent modulator output, as a function of the time delay $\tau$ between counter-propagating pulses, i.e. the oscillation between high and low output states and variation in oscillation magnitude with delay. Assuming that the two incident pulses have identical Gaussian temporal intensity profiles, one may readily derive from Eq. (1) that the normalized modulator output is a Gaussian function of delay with twice the width of the individual input pulse. Eq. (1) accurately reproduces the normalized output envelope of the experimental data in Fig. 3(a) with an envelope full-width half-maximum of 270 fs and values $A'_1 = 0.25$, and $A'_2 = 0.228$. [The small discrepancy between the sum $A'_1 + A'_2$ here and the average single-continuous-beam absorption levels for the two illumination directions (from Fig. 2) is attributed to the spectral width of the femtosecond pulses and to inhomogeneity across the metamaterial array (i.e. between precise locations on the sample at which the two types of measurement were performed).] Modulation contrast, defined as the ratio between upper and lower envelope limits and overlaid on the output signal data in Fig. 3(a), reaches a value of 2.8 at zero delay. Modulation bandwidth, even defined conservatively using the $1/e^2$ full-width of the envelope (459 fs), is 2.2 THz.

Closer detail of the output signal oscillation with mutual delay between pulses is shown in Fig. 3(b), for the ±60 fs range covered by the piezoelectric translation stage, and in Fig. 3(c), for the few cycles either side of the zero delay position. Within this limited range, where the oscillation magnitude depends very weakly on time delay, the pulses can be viewed as continuous waves at a wavelength equal to the pulse central wavelength (a good approximation because the 10 nm spectral width of the pulses is almost two orders of magnitude smaller than central wavelength) and Eq. (1) thereby reduces to,

$$A'_1[1 + \cos(\theta)] + A'_2 \qquad (2)$$

where $\theta$ is the relative phase difference between the two incident pulses ($\theta = 2\pi c\tau/\lambda$, where $c$ is the speed of light in vacuum and $\lambda$ is the center wavelength). This expression gives the analytical curve plotted alongside the experimental data in Fig. 3(c). The asymmetry in the latter is attributed to an imbalance between group velocity dispersion in the input beam paths, which prevents simultaneous phase-matching of all wavelength components.

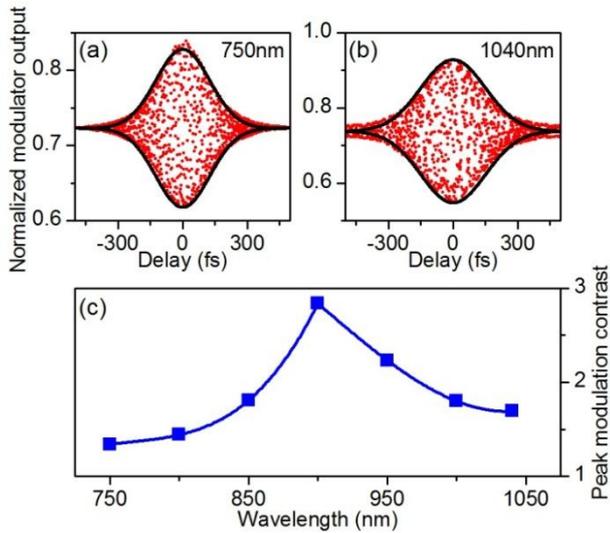

FIG. 4: Spectral dispersion of ultrafast coherent modulation contrast: (a, b) Representative experimental dependences of modulator output on the time delay between counter-propagating incident pulses at pulse central wavelengths of (a) (a) 750 and (b) 1040 nm. (c) Peak modulation contrast as a function of wavelength.

Experimental measurements such as shown in Figs. 3(a) and 3(b) were repeated for pulse central wavelengths of 750, 800, 850, 950, 1000, and 1040 nm. Representative plots (for 750 and 1040 nm) of modulator output against pulse delay are presented in Figs. 4(a) and 4(b), showing the characteristic modulation envelope observed in all cases. The dispersion of the peak modulator contrast ratio, i.e. the amplitude of this envelope, is shown in Fig. 4(c) and as one may expect found to reflect that of metamaterial absorption resonance centered at 900 nm (Fig. 2).

In summary, we have experimentally demonstrated an ultrafast all-optical modulator with functionality based on the femtosecond coherent control of absorption in a photonic metamaterial of nanoscale thickness. The underlying control mechanism is a linear interference effect and as such maybe implemented at arbitrarily low intensity, but it should be stressed that it is not based the splitting of light between output channels as in a conventional interferometer. Rather, the meta-device modulator is switched between high and low signal output states by manipulating the mutual phase and intensity of counter-propagating input pulses to select between high and low-absorption regimes of excitation at the metamaterial plane. Ultrafast modulator function is demonstrated here in the 750-1040 nm wavelength range of the experimental laser platform but the concept can be implemented freely across a broad visible to infrared wavelength band by varying the structural design of the metamaterial. Modulation bandwidth is limited in principle only by the spectral width of the metamaterial absorption resonance and as such may extend to terahertz frequencies. In practice, manufacturing imperfections, pulse duration and the matching of group velocity dispersion in the two beam paths will constrain achievable operating frequency and contrast. The peak contrast ratio of around 3:1 demonstrated here is already adequate for short-reach (intra-/interchip) optical interconnect applications in data processing architectures, where ratios of only 2.5:1 (modulation depths as low as 4 dB) can be sufficient.[25] Higher contrast still may be achieved in particular via pulse shaping.

Absorption is just one of many optical phenomena that may be efficiently controlled via the coherent interaction of optical waves on metamaterial nanostructures.[26] As such, in the coherent network environment, where meta-devices can be readily interconnected and cascaded, the coherent control paradigm may provide a family of solutions including logic gate functionality for ultrafast all-optical data processing.[27]

The authors thank Jianfa Zhang for helpful discussions. This work was supported by the Engineering and Physical Sciences Research Council [Project EP/G060363/1] (XF, JO, KFM, NIZ), The Royal Society and the Ministry of Education, Singapore [grant MOE2011-T3-1-005] (NIZ), and the National Science Council of Taiwan [Contracts NSC 102-2745-M-002-005-ASP, NSC101-2911-I-002-107] (MLT, DPT).